\documentclass[10pt]{iopart}

\usepackage{graphicx}

\usepackage{iopams}
\usepackage{amsfonts}

\begin{document}

\title[Inertia and Interaction]
{Physical properties of elementary particles: Inertia and Interaction}

\author{Mart\'{\i}n Rivas}
\address{Theoretical Physics Department, The University of the Basque Country (retired)\\ 
Bilbao, Spain}
\ead{martin.rivas@ehu.eus}

\begin{abstract}
Matter has two physical properties: Inertia and interaction. If we define the center of mass of an elementary particle in relation to its inertia, and a center of interaction in relation to its interactive properties, there are only two  possibilities to describe this elementary particle: that both points are the same or that they are different. If they are the same, what we describe is the point particle model, while if we consider them to be different, what we obtain is the description of an elementary spinning particle. If the center of interaction or center of charge is moving at the speed of light, completely determines also the dynamics of the center of mass, and when quantizing this model satisfies Dirac's equation. We obtain the classical description of the spinning Dirac particle. The general analysis of the interaction Lagrangian, suggests a modification of the minimal coupling Lagrangian, for a possible classical description of the strong and weak interaction.
\end{abstract}

\section{Introduction}
Material bodies have at least, two different physical properties: inertia and interaction.
Associated with inertia is the property we usually call mass, or inertial mass. For an arbitrary observer, the Principle of Inertia defines the concept of a free body. This definition is that a body is free for a certain observer if it is either at rest or moving with constant velocity. 
But the physical concept of velocity is mathematically understood to refer to the velocity of points, and in this particular case, to the velocity of a point, the center of mass (CM) of the body. 

A Principle of Relativity postulates the existence of a set of equivalent observers, and one of the criteria for equivalence is that if a body is free for one observer, it is free for all equivalent observers. In this way, we call the set of equivalent observers the class of inertial observers. The class of observers where the Inertia Principle holds.

The result is that inertial observers move with constant velocity relative to one another, and given a free body, 
it is always possible to find an inertial observer at rest with respect to it, i.e., with respect to its center of mass.

Regarding interaction, it is the physical property that matter has to relate to other matter, to cluster, to disperse, and to form more complex bodies. If matter had no ability to interact, it would mean that we could not relate to it, which in physical terms would imply that no interaction exists, and bodies are isolated from each other. No transfer of mechanical properties among bodies are allowed, including elastic scattering.

If ultimately all matter is made up of some final and indivisible elements that interact among each other and that we call elementary particles, we assume that these particles also possess the physical properties of inertia and interaction.

Since they are very small objects, we can accept that with respect to inertia there will exist a characteristic point called its center of mass (CM), and that with respect to its interaction properties there will exist at least one point that we will call the center of interaction or center of charge (CC). With this assumption we consider that elementary particles are massive particles.

There are only two possibilities: either both centers correspond to the same point, or they are two different points. It seems that theoretical physics has chosen as its basic elementary particle the first of these possibilities, ignoring the analysis that would arise from accepting the second.

If by interaction we mean electromagnetic interaction, then the center of interaction would be the center of electric charge. If we consider the electron to be an elementary particle, one of its measured properties is that it has angular momentum and magnetic moment, even at rest, and therefore this magnetic moment could well be produced by the motion of the center of charge with the center of mass at rest, and its angular momentum by the rotation of the particle.

This simple analysis rules out the first possibility, and therefore it seems more plausible that elementary matter would have two distinct centers, since inertia and interaction are two different physical properties.

This is precisely the description of an elementary spinning particle in the kinematical formalism developed in the book \cite{Rivasbook}. An updated version of some parts of the book and improved results, are contained in the Lecture Notes \cite{Lecture}.

We call a classical Dirac particle to the classical model that satisfies Dirac's equation when quantized \cite{Dirac}, and assume that leptons and quarks are Dirac particles. The main feature is that the classical Dirac particle is described by a single point ${\bi r}$, interpreted as the center of charge (CC) of the particle, moving at the speed of light, and that satisfies a system of ordinary differential equations of fourth order. These equations are obtained as the Euler-Lagrange equations of the mechanical system. One consequence of the Noether analysis is that a point ${\bi q}$, different than ${\bi r}$, can be defined and interpreted at the time coordinate $t$ in the inertial reference frame, as the center of mass of the particle (CM). In any inertial frame ${\bi r}$ and ${\bi q}$ are described simultaneously. In this formalism the CM is defined at every inertial reference frame at the same time than the CC, it is expressed in terms of the CC and their time derivatives and therefore their trajectories are computed and depicted simultaneously.
All observables of the classical Dirac particle, energy, linear momentum, angular momentum, angular velocity, magnetic and electric dipole moment, are expressed in terms of the point ${\bi r}$ and its different time derivatives \cite{pointelectron}. 

With the definition of the CM ${\bi q}$, the fourth-order differential equations of point ${\bi r}$ can be decoupled into a system of second-order differential equations for both points ${\bi r}$ and ${\bi q}$.

We know that in addition to the electromagnetic interaction, elementary particles are considered to interact at least in the form of strong, weak and gravitational interaction. The above analysis is perhaps suggesting that an elementary particle should have as many interacting centers as the different possible interactions. The kinematical formalism \cite{Rivasbook} is based on three independent fundamental principles: Restricted Relativity Principle, Variational Principle and Atomic Principle. It is the Atomic Principle that restricts the maximum number of classical variables that define the boundary variables of its Lagrangian description: The boundary variables manifold is necessarily a homogeneous space of the Poincar\'e group.

This restriction comes from the definition of a classical elementary particle. The requirement is that matter, when interacting, can produce scattering, bound systems, and if the interaction is of high energy, annihilation or creation of new particles. But if in the interacting process an elementary particle is not annihilated, its internal structure cannot be modified. This would justify the stability of matter at low energy or low temperature. An elementary particle cannot have excited states. If $x\equiv(x_1,\ldots,x_n)$ is the initial state of its Lagrangian description and this state changes to the state $y\equiv(y_1,\ldots,y_n)$ at a later time, if the internal structure has not been modified, it is possible to find another inertial observer at this time, that the state $y$ takes the same values of all variables like in the initial state $x$. Then there exists a Poincar\'e transformation $g$, such that $x=g*y$, or $y=g^{-1}*x$, for any pair of states, i.e., the boundary variables manifold of the Lagrangian evolution is a homogeneous space of the Poincar\'e group.

The maximum number of variables of a homogeneous space of the Poincar\'e group is $x\equiv(t,{\bi r},{\bi u},\balpha)$, interpreted as the time $t$, the position of a point ${\bi r}$, the velocity of this point ${\bi u}$ and the orientation $\balpha$ of a comoving orthogonal frame attached to the point ${\bi r}$. The essential three variables $\balpha$ are the way to characterize these three orthogonal unit vectors. Because the Lagrangian that describes this system must depend on the next order time derivative, it must be a function $L(t,{\bi r},{\bi u},\balpha,{\bi a},d\balpha/dt)$ also of the acceleration of the point ${\bi a}$, and of $d\balpha/dt$ or the angular velocity $\bomega$.  The dependence of the degrees of freedom $\balpha$ and its time derivative is through the dependence of the angular velocity $\bomega$,  which is a function of $\balpha$ and linear in $d\balpha/dt$. The free Lagrangian, being translation invariant, is a function $L_0({\bi u},{\bi a},\bomega)$ of the velocity and acceleration of the point and of the angular velocity of the comoving frame. 

This point ${\bi r}$ represents the center of charge CC, because when interacting, the interaction Lagrangian will be defined at this unique point, where the external fields are defined and the point ${\bi r}$, because the Lagrangian depends on the acceleration, satisfies a system of ordinary fourth-order differential equations. The model that satisfies Dirac equation when quantized is the model that $u=c$, and the point ${\bi r}$ is moving at the speed of light \cite{Dirac}. But we only have a single point ${\bi r}$, where we locate for the electron the charge $e$ of the particle. 

If we consider the boundary variables manifold given by $x\equiv(t,{\bi r}_1,{\bi r}_2)$ in terms of the position of two points, ${\bi r}_1$ and ${\bi r}_2$, two possible centers of two different interactions, then this manifold is not a homogeneous space of the Poincar\'e group. Given two sets of the above variables, a time translation transforms one time into the other. A space translation can transform one of the points, say ${\bi r}_1$, in its image, but it is impossible that the same space translation links the arbitrary initial and final positions of the other point ${\bi r}_2$. The classical elementary particle has to have a unique interaction center for all possible interactions. For quarks, the same point ${\bi r}$ represents the location of the center of electric charge, color charge and weak charge. For the classical description of leptons, the point ${\bi r}$ represents the location of the electric and weak charge. Concerning gravitation, if we identify inertial mass with gravitational mass, it seems to suggest that the CM is the location of the gravitational interaction center, but this has to be consistent with a general theory of gravitation of spinning particles.

%%%%%%%%%%%%%%%%%%%%%
\section{Constants of the motion of the free Dirac particle}
%%%%%%%%%%%%%%%%%%%%%
The free Lagrangian of the Dirac particle is a function $L_0({\bi u},{\bi a},\bomega)$ of the velocity and acceleration of the point ${\bi r}$, moving at the speed of light $u=c$, and of the angular velocity of the comoving frame. It is translation invariant and, therefore, independent of $t$ and ${\bi r}$. This  point satisfies a system of fourth order differential equations. The generalized coordinates are ${\bi r}$, ${\bi u}$ and $\balpha$, the boundary variables with the time excluded.
The dynamical equations of the ${\bi r}$ variables are:  \footnote{In this work, three-dimensional expressions like ${\bi a}={\partial L_0}/{\partial{\bi b}}$, have to be interpreted as $a_i={\partial L_0}/{\partial b_i}$, $i=1,2,3$.}
\[
\frac{\partial L_0}{\partial{\bi r}}-\frac{d}{dt}\left(\frac{\partial L_0}{\partial{\bi u}}\right)+\frac{d^2}{dt^2}\left(\frac{\partial L_0}{\partial{\bi a}}\right)=0,
\]
and because $L_0$ is space translation invariant, independent of ${\bi r}$, the first term cancels out and the above fourth-order differential equation is
\[
-\frac{d}{dt}\left[\frac{\partial L_0}{\partial{\bi u}}-\frac{d}{dt}\left(\frac{\partial L_0}{\partial{\bi a}}\right)\right]=-\frac{d{\bi p}}{dt}=0,
\]
and the conserved linear momentum, the canonical conjugate momentum of the ${\bi r}$ variables, is the term inside the squared brackets:
\begin{equation}
{\bi p}=\frac{\partial L_0}{\partial{\bi u}}-\frac{d}{dt}\left(\frac{\partial L_0}{\partial{\bi a}}\right).
\label{p}
\end{equation} 
The dynamical equations of the orientation variables $\balpha$, are:
\[
\frac{\partial L_0}{\partial{\balpha}}-\frac{d}{dt}\left(\frac{\partial L_0}{\partial{d\balpha/dt}}\right)=0,
\]
but the dependence of $L_0$ of $\balpha$ and $d\balpha/dt$ is through its dependence of the angular velocity $\bomega$, the above equation reduces to
\[
\frac{d{\bi W}}{dt}=\omega\times{\bi W}, \quad {\rm where}\quad {\bi W}=\frac{\partial L_0}{\partial{\bomega}},
\]
and ${\bi W}$ is the canonical conjugate momentum of the orientation variables ${\balpha}$. The canonical conjugate momentum of the variables ${\bi u}$, is finally,
\[
{\bi U}=\frac{\partial L_0}{\partial{\bi a}}.
\]
 The Hamiltonian is defined as usual as:
\begin{equation}
H={\bi p}\cdot{\bi u}+{\bi U}\cdot{\bi a}+{\bi W}\cdot\bomega-L_0,
\label{H}
\end{equation}
and is the conserved observable related to the time invariance of $L_0$. 

The invariance of $L_0$ under rotations defines the total angular momentum with respect to the origin of the observer frame as  \cite{Rivasbook}:
\[
{\bi J}={\bi r}\times{\bi p}+{\bi u}\times{\bi U}+{\bi W}={\bi L}+{\bi S},
\]
where ${\bi L}={\bi r}\times{\bi p}$ is the orbital angular momentum and 
\begin{equation}
{\bi S}={\bi u}\times{\bi U}+{\bi W}, 
\label{spin}
\end{equation}
is the spin with respect to the center of charge ${\bi r}$. Please remark that the spin is related to the dependence
of the Lagrangian of the acceleration ${\bi a}$, ${\bi U}={\partial L_0}/{\partial{\bi a}}$, and the angular velocity $\bomega$, ${\bi W}={\partial L_0}/{\partial{\bomega}}$.

Invariance of $L_0$ under pure Lorentz transformations defines the kinematical momentum with respect to the origin of the observer frame \cite{Rivasbook}:
\begin{equation}
{\bi K}=H{\bi r}/c^2-{\bi p}t-{\bi S}\times{\bi u}/c^2.
\label{K}
\end{equation}
If we define another position vector ${\bi k}$ by 
\begin{equation}
{\bi S}\times{\bi u}=H{\bi k}, \quad {\bi k}={\bi S}\times{\bi u}/H,
\label{k}
\end{equation}
the kinematical momentum can be rewritten as
\begin{equation}
{\bi K}=H{\bi q}/c^2-{\bi p}t,\quad {\rm where}\quad {\bi q}={\bi r}-{\bi k}={\bi r}-{\bi S}\times{\bi u}/H.
\label{q}
\end{equation}
Since $H$ and ${\bi p}$ are constants of the motion for the free particle, the time derivative of ${\bi K}$ leads to
${\bi p}=H{\bi v}/c^2$, where ${\bi v}=d{\bi q}/dt$ is the constant velocity of the point ${\bi q}$. This point can be interpreted as the center of mass (CM) of the Dirac particle. Because the invariant and constant of the motion
$H^2-{\bi p}^2c^2=m^2c^4$ defines the mass of the particle, we arrive to 
\begin{equation}
{\bi p}=\gamma(v)m{\bi v},\quad H=\gamma(v)mc^2,\quad \gamma(v)=(1-v^2/c^2)^{-1/2},
\label{Hpg}
\end{equation}
and linear momentum and energy are expressed in terms of the CM velocity, like in the case of the point particle. It is shown in \cite{Rivasbook} that ${\bi W}$ is orthogonal to the CC velocity ${\bi u}$ and thus the spin ${\bi S}$ in (\ref{spin}) is also orthogonal to ${\bi u}$. The cross product with ${\bi u}$ in (\ref{k})
\begin{equation}
({\bi S}\times{\bi u})\times{\bi u}=H{\bi k}\times{\bi u},\quad\rightarrow\quad -{\bi S}c^2=H({\bi r}-{\bi q})\times{\bi u},
\label{S}
\end{equation}
and by making use of (\ref{Hpg}) the spin with respect to the CC is written as
\begin{equation}
{\bi S}=-\gamma(v)m({\bi r}-{\bi q})\times{\bi u}.
\label{SCC}
\end{equation}
It is expressed in terms of the separation of the two points ${\bi r}$ and ${\bi q}$, and of the velocities of both points.

The spin with respect to the CM ${\bi q}$, is defined at the same time, as usual
\begin{equation}
{\bi S}_{CM}={\bi S}+({\bi r}-{\bi q})\times{\bi p}=-\gamma(v)m({\bi r}-{\bi q})\times({\bi u}-{\bi v}).
\label{SCM}
\end{equation}
It also depends on the separation of the two points ${\bi r}$ and ${\bi q}$, and of the velocities of both points.

\begin{figure}[!hbtp]\centering%
\includegraphics[width=5cm]{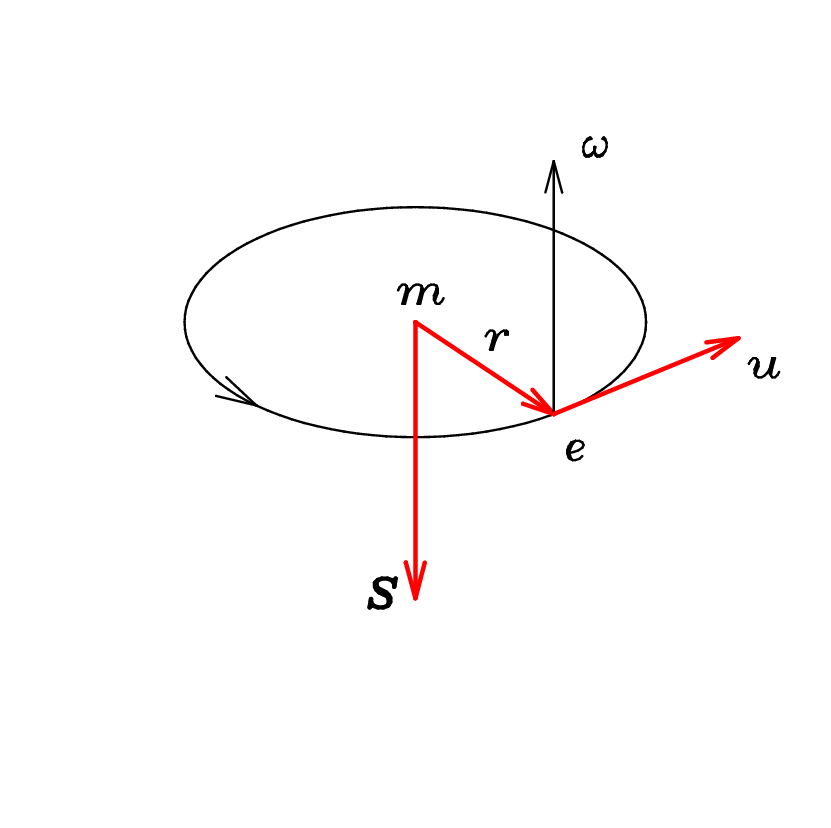}
\caption{Motion of the CC of the electron ${\bi r}$ in the center of mass frame given in (\ref{espinCM}). The CM is located at the origin in this frame. The angular velocity $\bomega$, because ${\bi v}=0$ has no component along the velocity ${\bi u}$, the trajectory is flat and has no torsion. If the CM velocity ${\bi v}\neq0$ the angular velocity has a component along ${\bi u}$, that produces the torsion of the trajectory of the CC. The spin in the center of mass frame has the opposite direction to the angular velocity $\bomega$. Remark that the orientation of the spin with respect to the plane of the trajectory of point ${\bi r}$, is opposite to the orientation of the angular momentum if the trajectory were the motion of a point mass.} 
\label{fig1}
\end{figure}

They satisfy two different differential equations. The time derivatives of (\ref{SCC}) and (\ref{SCM}) are, respectively,
\begin{equation}
\frac{d{\bi S}}{dt}={\bi p}\times{\bi u},\quad \frac{d{\bi S}_{CM}}{dt}=0.
\label{espindyn}
\end{equation}
The CM spin is conserved but the spin with respect to the CC satisfies the same dynamical equation than Dirac's spin operator in the quantum case. It is the classical spin equivalent to Dirac's spin operator. In the center of mass frame (${\bi q}={\bi v}=0$) and both spins take the same value. In the center of mass frame and for a free electron, both spins are constant, and point ${\bi r}$ satisfies the differential equation
\begin{equation}
 {\bi S}=-m{\bi r}\times{\bi u},
\label{espinCM}
\end{equation}
so that the point ${\bi r}$ describes a circle of constant radius $R_0=S/mc$ at the speed of light in a plane orthogonal to the constant spin, as depicted in figure {\bf\ref{fig1}}.

The time derivative of the constant kinematical momentum (\ref{K}) gives
\begin{equation}
H{\bi u}-c^2{\bi p}-\frac{d{\bi S}}{dt}\times{\bi u}-{\bi S}\times\frac{d{\bi u}}{dt}=0.
\label{KHDir}
\end{equation}
Since from (\ref{espindyn}) we substitute $d{\bi S}/dt={\bi p}\times{\bi u}$, taking the cross product of the resulting expression with the acceleration $d{\bi u}/dt$ and from the definition of ${\bi q}={\bi r}-{\bi k}$ in (\ref{q}), and the expressions of $H$ and ${\bi p}$ of (\ref{Hpg}) we get that the CM position can be written as
\begin{equation}
{\bi q}={\bi r}+\frac{c^2-{\bi v}\cdot{\bi u}}{a^2}{\bi a}.
\label{CMpos}
\end{equation}
The separation vector between the CM and CC, ${\bi q}-{\bi r}$, has the direction of the CC acceleration ${\bi a}$ and is orthogonal to the CC velocity ${\bi u}$. The CC performs a central motion at the constant velocity $c$, around the CM. All variables in the expression (\ref{CMpos}) are considered at the same time $t$, the time coordinate of the corresponding inertial observer.

%%%%%%%%%%%%%%%%%%%
\subsection{Intrinsic properties of the Dirac particle}
%%%%%%%%%%%%%%%%%%%
The Classical Dirac particle has two invariant properties related to the two Casimir operators of the Poincar\'e group, namely the invariant related to the four-momentum $p^\mu\equiv(H/c,{\bi p})$, $p^\mu p_\mu=m^2c^2$, and the invariant related to the Pauli-Lubanski four-vector $w^\mu=(1/2)\epsilon^{\mu\nu\sigma\lambda}p_\nu J_{\sigma\lambda}$, that can be written in terms of $H$, ${\bi p}$ and ${\bi S}_{CM}$ as:
\begin{equation}
w^\mu\equiv({\bi p}\cdot{\bi S}_{CM},H{\bi S}_{CM}/c),\quad w^\mu w_\mu=-m^2c^2 S(0)^2,
\label{wmu}
\end{equation}
so that the two parameters, the mass $m$ and the absolute value of the spin in the center of mass frame $S(0)$, are the intrinsic properties of the electron.

The observables $H$, ${\bi p}$, the two spins ${\bi S}$, ${\bi S}_{CM}$, the synchronous location of the CM ${\bi q}$, can be expressed in terms of the point ${\bi r}$ and their time derivatives, simultaneously. 
%%%%%%%%%%%%%%%%%%%
\subsection{Dirac's Hamiltonian}
%%%%%%%%%%%%%%%%%%%
If in the above equation (\ref{KHDir}) we make the scalar product with ${\bi u}$ we get:
\begin{equation}
H={\bi p}\cdot{\bi u}+\frac{1}{c^2}{\bi S}\cdot\left({\bi a}\times{\bi u}\right),
\label{Hdir}
\end{equation}
This linear expression in terms of $H$ and ${\bi p}$, represents the classical equivalent of Dirac's Hamiltonian. The energy of the Dirac particle has two terms: the first ${\bi p}\cdot{\bi u}$, represents the translation energy, depends on the mass and is related to the inertia of the particle. The second, ${\bi S}\cdot\left({\bi a}\times{\bi u}\right)/c^2$, is positive definite and non-vanishing, represents the rotation energy and depends on the other intrinsic property of the Dirac particle, the spin. Because ${\bi p}$ is the canonical conjugate momentum of the variable ${\bi r}$, we obtain the Hamilton equation $\partial H/\partial{\bi p}={\bi u}$.

%%%%%%%%%%%%%%%%%%%%%
\section{The interaction Lagrangian}
\label{inter}
%%%%%%%%%%%%%%%%%%%%%
The Atomic Principle requires that any interaction that does not annihilate the Dirac particle cannot modify its internal structure, and thus, mass an spin must remain invariant. In the definition of spin (\ref{spin}) if the interaction does not modify this definition, the interaction Lagrangian cannot depend on the acceleration ${\bi a}$ and the angular velocity $\bomega$, so that the most general interaction Lagrangian will be a function of the variables $L_I(t,{\bi r},{\bi u})$, independent of ${\bi a}$ and $\bomega$.

If instead of using the time evolution description in terms of the time variable of a particular inertial frame, we make the evolution in terms of some arbitrary invariant parameter $\tau$, the same for all inertial observers, the formalism is independent of the particular time of the observers. Thus the action integral is rewritten as
\[
\int_{t_1}^{t_2}\,L dt=\int_{\tau_1}^{\tau_2}\,L (dt/d\tau) d\tau=\int_{\tau_1}^{\tau_2}\widetilde{L}d\tau,
\]
where $\widetilde{L}=L\dot{t}$, and the dot means the derivative with respect to $\tau$. In terms of derivatives with respect to $\tau$, the Lagrangian $L_I$ is a homogeneous function of zero degree in terms of $\dot{t}$ and $\dot{\bi r}$,
\[
L_I(t,{\bi r},{\bi u})\equiv L_I(t,{\bi r},\dot{\bi r}/\dot{t}),
\]
because the velocity is expressed as ${\bi u}=d{\bi r}/dt=\dot{\bi r}/\dot{t}$. It is also $\tau$-independent. Therefore $\widetilde{L}_I=L_I\dot{t}$, is a function of $\widetilde{L}_I(t,{\bi r},\dot{t},\dot{\bi r})$, $\tau-$independent homogeneous function of first degree in terms of $\dot{t}$ and $\dot{\bi r}$. Euler's theorem implies that
\[
\widetilde{L}_I=\frac{\partial\widetilde{L}_I}{\partial{\dot{t}}}\dot{t}+\frac{\partial\widetilde{L}_I}{\partial{\dot{\bi r}}}\cdot\dot{\bi r}=A_0(t,{\bi r},{\bi u})\dot{t}+{\bi A}(t,{\bi r},{\bi u})\cdot\dot{\bi r},
\]
in terms of a scalar and vector functions $A_0(t,{\bi r},{\bi u})$ and ${\bi A}(t,{\bi r},{\bi u})$, respectively,
that are homogeneous functions of zero degree in terms or $\dot{t}$ and $\dot{\bi r}$, and thus functions of $(t,{\bi r},{\bi u})$.
The original Lagrangian in the time evolution description in any frame $L_I=\widetilde{L}_I/\dot{t}$, can always be written in terms of these four arbitrary functions $A_\mu$, $\mu=0,1,2,3$.
\begin{equation}
L_I=A_0(t,{\bi r},{\bi u})+{\bi u}\cdot{\bi A}(t,{\bi r},{\bi u}).
\label{LI}
\end{equation}
Since
\begin{equation}
A_0=\frac{\partial\widetilde{L}_I}{\partial{\dot{t}}},\quad A_j=\frac{\partial\widetilde{L}_I}{\partial{\dot{r}_j}},\quad {\rightarrow}\quad  \frac{\partial A_0}{\partial{\dot{r}_j}}=\frac{\partial A_j}{\partial{\dot{t}}},
\label{a0ak}
\end{equation}
and these four functions are not independent functions.

The function
\begin{equation}
A_j=\frac{\partial\widetilde{L}_I}{\partial{\dot{r}_j}}=\frac{\partial(\dot{t}{L}_I)}{\partial{\dot{r}_j}}=\dot{t}\frac{\partial{L}_I}{\partial{\dot{r}_j}}=\dot{t}\frac{\partial{L}_I}{\partial{{u}_k}}\frac{\partial{u}_k}{\partial{\dot{r}_j}}=\dot{t}\frac{\partial{L}_I}{\partial{{u}_k}}\frac{1}{\dot{t}}\delta_{kj}=\frac{\partial{L}_I}{\partial{{u}_j}},
\label{deLA3}
\end{equation}
and the vector function ${\bi A}=\partial{L}_I/\partial{\bi u}$, is the partial derivative of the temporal Lagrangian $L_I$ with respect to ${\bi u}$.
For $A_0$ we make:
\[
A_0=\frac{\partial\widetilde{L}_I}{\partial{\dot{t}}}=\frac{\partial(\dot{t}{L}_I)}{\partial{\dot{t}}}=L_I+\dot{t}\frac{\partial{L}_I}{\partial{\dot{t}}}=L_I+\dot{t}\frac{\partial{L}_I}{\partial{{u}_k}}\frac{\partial{u}_k}{\partial{\dot{t}}}=
\]
\[
=L_I+\dot{t}\frac{\partial{L}_I}{\partial{{u}_k}}\left(\frac{-\dot{r_k}}{\dot{t}^2}\right)=L_I-{\bi A}\cdot{\bi u},
\]
that is (\ref{LI}). But from this
\begin{equation}
\frac{\partial A_0}{\partial u_j}=\frac{\partial L_I}{\partial u_j}-A_j-\frac{\partial A_k}{\partial u_j}u_k=-\frac{\partial A_k}{\partial u_j}u_k. 
\label{a0uk1}
\end{equation}
Form (\ref{a0ak}) we have:
\[
 \frac{\partial A_0}{\partial{\dot{r}_j}}= \frac{\partial A_0}{\partial{{u}_k}}\frac{\partial{u}_k}{\partial{\dot{r}_j}}=\frac{1}{\dot{t}} \frac{\partial A_0}{\partial{{u}_j}}=\frac{\partial A_j}{\partial{\dot{t}}}=\frac{\partial A_j}{\partial{u_k}}\frac{\partial{u}_k}{\partial\dot{t}}=\frac{\partial A_j}{\partial{u_k}}\frac{-\dot{r}_k}{\dot{t}^2}.
\]
We also get that
\begin{equation}
\frac{\partial A_0}{\partial u_j}=-\frac{\partial A_j}{\partial u_k}u_k.
\label{a0uk2}
\end{equation}
If it happens that ${\bi A}$ is independent of ${\bi u}$, $A_0$ is also independent of ${\bi u}$. This is what we get if we assume that the interaction Lagrangian describes the so-called {\it minimal coupling}, where $-A_0(t,{\bi r})$
is $e$ times the scalar potential and ${\bi A}(t,{\bi r})$ is $e$ times the vector potential of the external electromagnetic field, and $e$ the value of the electric charge of the particle. But this analysis suggests that more general classical interactions than the electromagnetic interaction can be analyzed, provided we select the functions $A_\mu$, functions of the velocity of the CC ${\bi u}$. The minimal coupling interaction Lagrangian is linear in the CC velocity, and the functions $A_\mu$ are velocity independent.

From $L_I(t,{\bi r},{\bi u})$, and (\ref{deLA3}) we get $\partial L_I/\partial{\bi u}={\bi A}$. A first integral gives $L_I={\bi A}\cdot{\bi u}+B(t,{\bi r})$, where $B$ is an arbitrary scalar function of $(t,{\bi r})$, and therefore $A_0$ is a function independent of the velocity ${\bi u}$. From (\ref{a0uk1}) and (\ref{a0uk2}) we obtain that the possible dependence of the vector function ${\bi A}$ of the velocity variables, must satisfy
\begin{equation}
\left(\frac{\partial A_j}{\partial u_k}-\frac{\partial A_k}{\partial u_j}\right)u_k=0, \quad j=1,2,3.
\label{ajk}
\end{equation}
As an example, the linear functions of ${\bi u}$, ${\bi A}=f(t,{\bi r}){\bi u}$ and  ${\bi A}={\bf g}(t,{\bi r})\times{\bi u}$, satisfy (\ref{ajk}), but since $u^2=c^2$, the interaction Lagrangian reduces to the scalar function $L_I=V(t,{\bi r})$, independent of ${\bi u}$. But in principle, we can look for interaction Lagrangians more general than the minimal coupling to describe other classical interactions.

Since $A_0$ is independent of ${\bi u}$, from (\ref{a0uk2}) we obtain $({\partial A_j}/{\partial u_k})u_k=0$, so that ${\bi A}$ is a homogeneous function of ${\bi u}$ of zero degree. It is a function of ${\bi u}/u$, and thus this possible interaction Lagrangian must depend, in addition to the time $t$ and position ${\bi r}$, on the two compact variables, the zenithal angle $\theta$ and the azimuthal angle $\phi$, of the non-vanishing CC velocity ${\bi u}$. This more general interaction Lagrangian must be dependent of the orientation of the CC velocity of the particle.

%%%%%%%%%%%%%%%%%%%
\section{Dynamical equations}
%%%%%%%%%%%%%%%%%%%
The dynamical equations of the center of mass ${\bi q}$ and the center of charge ${\bi r}$ of the Dirac particle under any electromagnetic field are obtained from the general Lagrangian $L=L_0+L_{em}$, where $L_0({\bi u},{\bi a},\bomega)$ is the free Lagrangian. The interaction Lagrangian $L_{em}=-eA_0(t,{\bi r})+e{\bi u}\cdot{\bi A}(t,{\bi r})$, is written in terms of the electric charge $e$ and the scalar and vector potentials $A_0$ and ${\bi A}$ defined at time $t$ at the center of charge position ${\bi r}$. Euler-Lagrange equations give rise to the first order system:
\begin{eqnarray}
\frac{d{\bi q}}{dt}&=&{\bi v},\quad\frac{d{\bi v}}{dt}=\frac{1}{m\gamma(v)}\left[{\bi F}-\frac{\bi v}{c^2}\left({\bi F}
\cdot{\bi v}\right)\right],\label{eq:d2qdt2}\\
\frac{d{\bi r}}{dt}&=&{\bi u},\quad \frac{d{\bi u}}{dt}=\frac{c^2-{\bi v}\cdot{\bi u}}{({\bi q}-{\bi r})^2}({\bi q}-{\bi r}),\label{eq:d2rdt2}
 \end{eqnarray}
where the external force ${\bi F}=e{\bi E}(t,{\bi r})+e{\bi u}\times{\bi B}(t,{\bi r})$, is the Lorentz force defined at the CC, and $m$ and $e$ are the mass and the electric charge of the electron, respectively. The two velocities satisfy the constraints $u=c$, $v<c$.  Equation (\ref{eq:d2qdt2}) is just the Euler-Lagrange equation of the ${\bi r}$ variables $d{\bi p}/dt={\bi F}$, after taking the time derivative of ${\bi p}=\gamma(v)m{\bi v}$, and leaving on the left-hand side the CM acceleration $d{\bi v}/dt$. The equation (\ref{eq:d2rdt2}) is equation (\ref{CMpos}) when writing on the left-hand side the CC acceleration ${\bi a}$.
The original fourth-order differential equations for the point ${\bi r}$ have been rewritten as a second-order system of differential equations for both points ${\bi q}$ and ${\bi r}$, or as the first order system (\ref{eq:d2qdt2}) and (\ref{eq:d2rdt2}) for the variables ${\bi q}$, ${\bi v}$, ${\bi r}$ and ${\bi u}$, more suitable for a numerical integration, without specifying the Poincar\'e invariant free Lagrangian $L_0$, and with the minimal coupling interaction Lagrangian $L_{em}$.

In the publication \cite{ClassicalDiracI} we analyze the interaction of the Dirac particle with external uniform and oscillating electric and magnetic fields, we make a link to some Mathematica notebooks where the differential equations are solved numerically, and also to Wolfram's Community groups \cite{electric} where some notebooks are included.
In the work \cite{ClassicalDiracII} we describe the interaction of the classical Dirac particle with a monochromatic circularly polarized plane wave, and in \cite{2elec} we describe the interaction of two Dirac particles, where the Poincar\'e invariant interaction Lagrangian is velocity dependent, it is not of a minimal coupling structure but is a kind of instantaneous action at a distance of the two particles, and depends on the relative orientations of the CC velocities of both particles, as suggested in Section {\bf\ref{inter}} when describing the general interaction Lagrangian. Because electrons are leptons, a possible classical weak interaction description of leptons, should have a weak interaction Lagrangian that depends on the orientation of the CC velocity.

%%%%%%%%%%%%%%%%%%%
\section{Radiation Reaction}
%%%%%%%%%%%%%%%%%%%

We have considered that elementary particles have inertia and therefore they are massive particles. The main characteristic is that they have a CC and CM as two different points. In \cite{Radreaction} we have shown that in the case of the electromagnetic interaction the radiation occurs because the energy expended by the field is the work of the external Lorentz force along the CC trajectory, while the change of the mechanical energy of the particle is the work along the CM trajectory. If these two works are different, energy conservation of the closed system particle+field, implies that there is a continuous transfer of electromagnetic energy, linear momentum and angular momentum between the particle and the field. Preliminary calculations of the transfer of energy and linear momentum, show that the amount of energy equals the value of the linear momentum transfer and this can be interpreted as if a massless electromagnetic quantum of energy and linear momentum has been transferred to the field. To consider that this transfer corresponds to a photon we have to compute the transfer of angular momentum. Because the transfer is continuous, this will correspond to a photon if after some time $\Delta t$, the total continuous transfer of angular momentum amounts $\hbar$, the angular momentum of a photon. 

The requirement that the value of the spin for the center of mass observer has led to modify the dynamical equation (\ref{eq:d2qdt2}) $d{\bi p}/dt={\bi F}$, by introducing a breaking or accelerating term along the CM velocity ${\bi v}$,
\[
\frac{d{\bi p}}{dt}={\bi F}-\gamma^2\left[({\bi u}-{\bi v})\cdot{\bi F}\right]{\bi v},
\]
term that is related to the difference of the work per unit time, of the external Lorentz force ${\bi F}$ along the CC and CM trajectories. 

An accelerated electron by an external Lorentz force, needs some time $\Delta t$, to interchange energy linear momentum and also a unit $\hbar$ of angular momentum with the external field.
If this happens to be correct, the interaction between the field and the particle corresponds to the creation of a quantum of electromagnetic energy, linear momentum and of angular momentum 1, in natural units. With the above consideration that elementary particles are massive particles, the created photon is not an elementary particle. We cannot define its center of mass, but corresponds to an uncharged massless object that moves at the speed of light along a straight line. The electromagnetic interaction implies the assumed interchange of massless bosons of spin 1, as derived from Quantum Electrodynamics, that can be created and annihilated during the process of interaction. The photon will be the result of the electromagnetic interaction.

What follows is a conjecture. Let us assume that with the freedom we have to determine another possible interaction Lagrangian, we were able to obtain a classical description of other interactions. The CC will be the same interacting point as the center of electric charge of the Dirac particle, and being different than the CM, there will be a difference between the external work and the increase of the mechanical energy of the particle for the new interaction. If it happens that this interchange of mechanical properties between the particle and the field produces new particles, they will be bosons because the angular momentum of the Dirac particle has to remain the same. If they are massless, they must be moving at the speed of light and must be bosons because the spin of leptons and quarks is $1/2$. 

We have to obtain first a new non-minimal coupling interaction Lagrangian, that predicts a short-range interacting force, but the results of quantum chromodynamics and weak interactions predict, in the quantum analysis, the existence of intermediate bosons, massless gluons in the strong interaction and very massive bosons for the weak interaction. The existence of massive bosons has also been considered in the general formalism \cite{Rivasbook}.
Spinning particles with a CC different than the CM, such that the CC is moving faster than light can also be described in this formalism. The CM is moving at a velocity $v<c$, but when quantizing these models, they only produce integer spin particles. Faster than light motions of the CC of an elementary particle only quantize with integer values. If this happens to be correct, the $Z^0$ and $W^{\pm}$ massive bosons of the weak interaction, will correspond, from the classical point of view, to massive particles with a CM that can be at rest, but whose CC is moving faster than light.

%%%%%%%%%%%%%%%%%%%%%%%%%%%% 


\section*{References}
\begin{thebibliography}{10}
  \bibitem{Rivasbook}{M. Rivas (2001),
 {\it Kinematical Theory of Spinning Particles}, Fundamental Theories of Physics Series, Vol 116, Dordrecht, Holland.\\{https://link.springer.com/book/10.1007/0-306-47133-7}}
\bibitem{Lecture}{M. Rivas (2024), {\it Kinematical theory of elementary spinning particles}, Lecture Notes Bilbao 2024, {https://www.researchgate.net/publication/381660537}.\\ For Spanish speaking readers there is a Spanish version at\\
{https://www.researchgate.net/publication/381660817}} 
\bibitem{Dirac}{M. Rivas (1994), {\it Quantization of generalized spinning particles: New derivation of Dirac’s equation}, J. Math.Phys. {\bf 35}, 3380-3399 .}
\bibitem{pointelectron}{M. Rivas (2026), {\it Is it possible to describe an electron by the evolution of a single point?}
preprint https://arxiv.org/abs/2601.11597v3.}
 \bibitem{ClassicalDiracI}{J. Barandiaran and M. Rivas (2025), {\it Classical Dirac Particle I}, 
J. Theor. Phys. and Math. Res. {\bf 3} 1-26, https://dx.doi.org/10.64030/3065-8802.03.02.04\\
preprint https://arxiv.org/abs/2505.08794v3.} 
\bibitem{electric}{J. Barandiaran and M. Rivas (2025), {\it Dirac Particle in a uniform and oscillating Magnetic and Electric field}\\{ https://www.zitter-institute.org/p/dirac-particle-magnetic-electric-field.html}}\\
The Wolfram Community Project has included the following two notebooks:\\J. Barandiaran (2025) 
{\it Classical Dirac Particle I}\\ https://community.wolfram.com/groups/-/m/t/3521996\\
{\it Classical Dirac particle I: the electron spinning particle model for dynamic simulations (Part II)}
 https://community.wolfram.com/groups/-/m/t/3543405
 \bibitem{ClassicalDiracII}{J. Barandiaran and M. Rivas (2025), {\it Classical Dirac Particle II. Interaction with an electromagnetic plane wave}, 
preprint https://arxiv.org/abs/2508.16649v1.} 
\bibitem{2elec}{J. Barandiaran and M. Rivas (2025) {\it Poincar\'e invariant interaction of two Dirac particles}, preprint https://arxiv.org/abs/2501.10445v2.}
\bibitem{Radreaction}{M. Rivas (2026), {\it Classical Dirac particle: Mass and Spin invariance and radiation reaction},
preprint https://arxiv.org/abs/2512.10505v2.}

\end{thebibliography}
\end{document}